\documentclass[%
 reprint,
 amsmath,amssymb,
 aps,
]{revtex4-2}
\usepackage{graphicx}
\usepackage{dcolumn}
\usepackage{bm}
\usepackage{braket}
\usepackage{float}
\usepackage{multirow}
\allowdisplaybreaks[1]

\def\leqn#1{(\ref{#1})}

\def\beq{\begin{equation}}
\def\eeq#1{\label{#1}\end{equation}}
\def\eeqn{\end{equation}}

\newenvironment{Eqnarray}%
   {\arraycolsep 0.14em\begin{eqnarray}}{\end{eqnarray}}
\def\beqa{\begin{Eqnarray}}
\def\eeqa#1{\label{#1}\end{Eqnarray}}
\def\eeqan{\end{Eqnarray}}
\def\CR{\nonumber \\ }

\def\leqn#1{(\ref{#1})}

\def\ee{e^+e^-}
\def\half{{1\over 2}}
\def\One{{\mathbf 1}}

\begin{document}

\title{The Sensitivity of Higgs Factories to Composite Higgs Models\\  via Precision Measurements }

\author{Kamal Maayergi, Devin G.~E.~Walker}
\affiliation{Department of Physics and Astronomy, Dartmouth College \\Hanover, NH 03755 USA}%

\author{Ora Cullen}
\affiliation{%
 Department of Physics and Astronomy, Dartmouth College \\Hanover, NH 03755 USA
}%
\affiliation{School of Mathematics, Trinity College Dublin, Ireland}

\author{Michael E.~Peskin}
\affiliation{
SLAC National Accelerator Laboratory\\ Stanford University\\ Menlo Park, California 94025 USA
}%

\begin{abstract}
We investigate the potential of precision Higgs factory measurements to discover signatures of a representative model of electroweak symmetry breaking in which the Higgs boson arises as a composite Nambu-Goldstone boson.  In this model, as in other models of the ``Little Higgs" or Natural Composite Higgs type, the primary perturbations of the Standard Model come from effects of vectorlike top quark partners. We carry out an explicit calculation of the Higgs potential in this model. Applying phenomenological constraints, we are left with a 3-dimensional parameter space.  We then present results from a complete scan of this parameter space. The region in which significant departures from the Standard Model predictions extends to models in which the lightest top quark partner has a mass above 3~TeV.  Little Higgs models with such heavy top partners also predict significant deviations from the Standard Model in the top quark electroweak couplings, in particular, in the model studied here, in the $t_L$ coupling to the $Z$ boson.
\end{abstract}

\maketitle
\section{\label{sec:Intro} Introduction}

The Higgs boson was discovered in 2012
\cite{ATLAS:2012yve,CMS:2012qbp} at the Large Hadron Collider (LHC).
Since then, the two main experiments focused on probing the couplings
of the Higgs to vector bosons and fermions have been ATLAS and CMS [3,
4]. The results to date have all been consistent with the predictions
for the unique Higgs boson of the Standard Model (SM), with coupling
measurements achieving accuracies at the 10\%–20\% level for the $W$
and $Z$ bosons and the third-generation quarks and leptons. The
consistency of these results with the theoretical predictions has
firmly established the 125~GeV Higgs boson as the main source of
spontaneous electroweak symmetry breaking. The High-Luminosity run of
the LHC (HL-LHC) is an upgrade, currently underway,
that will
sharpen this picture by increasing the precision of coupling
measurements to the 1–4\% level; and extending the
search for additional fermions beyond the 1~TeV mass scale~\cite{ATLAS:2022hsp}.

Despite the great success of the LHC and its findings, questions
still remain in our understanding, namely, the nature of the Higgs
boson and the mechanism of electroweak symmetry breaking. In the SM,
electroweak symmetry breaking is an assumption, implemented by the
choice of input parameters. Moreover, the SM does not explain the
large hierarchy of fermion masses, with ratios exceeding 100,000
between the heaviest and lightest fermions. It is
therefore crucial to obtain new insights into the properties and
possible partners of the 125~GeV Higgs boson. This motivation has led
to proposals for new $\ee$ colliders designed as ``Higgs factories''
to enable higher-precision measurements of the known Higgs boson
$h$~\cite{Dawson:2022zbb}. The hope is that precision measurements on
the Higgs boson can give insight into its possible partners and  
new interactions.  The
top quark often plays a role in models of electroweak symmetry
breaking, and so precision studies of the top quark are also part of
this story.   This paper will concentrate on a class of models of
electroweak symmetry breaking in which the top quark plays an
essential role.

Current proposals for Higgs factories now under consideration
include the ILC~\cite{ILCInternationalDevelopmentTeam:2022izu} in
Japan, the CEPC in China~\cite{CEPCPhysicsStudyGroup:2022uwl}, and the
FCC-ee and LCF at
CERN~\cite{FCC:2025lpp,LinearColliderVision:2025hlt}. The projected
precision for Higgs boson coupling
measurements at these facilities is quite similar; in this paper, we
refer to the LCF projections for definiteness. The LCF projections are
based on full-simulation studies using the ILD detector model, with
all relevant physics and beam-related backgrounds included. See
\cite{ILDConceptGroup:2020sfq,Ono:2012oyw}
for more details.

 A question that is often asked about precision measurements is: What
 is the mass reach for the sensitivity to new particles?  A simple
 estimate of the sensitivity can be derived from the idea that
 corrections to the Standard Model (SM) due to new heavy particles
 should be parametrized by higher-dimension operators in Standard
 Model Effective Field Theory (SMEFT). The leading such operators are
 at
 dimension~6, leading to an estimate of the size of the corrections as
\beq
v^2/M^2 \sim  1\% \ ,
\eeq{corrval}
where $v$ is the Higgs field vacuum expectation value and $M$ is the
heavy particle mass, possibly suppressed further by a factor of
$\alpha_w$.  However, \leqn{corrval} is only an order-of-magnitude
estimate which might be modified by a large dimensionless prefactor.
There are certainly beyond-SM theories that give small corrections to
the Higgs couplings.  A more meaningful question  is
whether there are opportunities for discovery with Higgs
factory precision measurements~\cite{Peskin:2022pfv}.  To address
 this, it is necessary to study explicit models of new
physics. This paper
follows the approach of our earlier paper~\cite{Maayergi:2025ybi},
applied now to a very different class of models.

It is well-known that Effective Field Theory descriptions of
composite models of the Higgs boson
such as the Strongly-Interacting Light Higgs
(SILH)~\cite{Giudice:2007fh} can predict observable deviations in the
Higgs couplings originating from physics at mass scales of 10~TeV and
above.
In this paper,
we will address the mass reach of precision measurements
in more explicit models of the Higgs
boson in a composite
state.

Observations at the
LHC indicate that the 125~GeV Higgs boson has properties similar to
those of an elementary scalar particle.   A mechanism that  makes this
natural in the context of composite Higgs bosons
is for the Higgs doublet fields to be a set of Nambu-Goldstone bosons
associated with a strong interaction symmetry breaking at a high
energy scale~\cite{Kaplan:1983fs,Kaplan:1983sm}.
Some explicit models that implement this idea are the 
Little Higgs models~\cite{Schmaltz:2005ky} or Natural Composite Higgs models~\cite{Katz:2005au,Nelson:2018iuc}.  In models of Higgs as a
Nambu-Goldstone boson, the
high-energy theory has a large symmetry group $G$ broken to a subgroup
$H$ by strong interaction effects. $H$ would contain the electroweak
$SU(2)\times U(1)$ as a subgroup.   If $H$ remained exact, the Higgs
fields would be massless.  However, the original theory could also include
additional interactions that break $G$ explicitly.  Couplings to the
electroweak interactions could provide some of these effects.   It is
still a problem if the perturbations induce quadratic divergences in
the Higgs in one-loop diagrams.  Little Higgs models avoid this issue
using a mechanism called
``collective symmetry breaking''~\cite{Arkani-Hamed:2002sdy}.   
In models implementing this mechanism, each individual weak coupling
preserves enough
symmetry that the Higgs is an exact Nambu-Goldstone boson and is forbidden
from getting  a
mass.  This happens, for example, in models of chiral symmetry
breaking in which the left- and right-handed strongly interacting
fermions have different weak  interactions.
It is not until two or more couplings are turned on simultaneously
that the symmetry is broken sufficiently that the Higgs field  acquires
a mass and becomes a pseudo-Nambu-Goldstone boson
(pNGB).  Contributions to the Higgs mass involve these couplings
collectively, and so the Higgs potential and the associated
symmetry-breaking minimum appears in the second order of the
perturbation expansion.  This is an example of a scenario in which an
underlying symmetry-breaking at 10~TeV can feed down in two stages to
a Higgs mass at the scale of 100~GeV.

In these models, signs of the physics of electroweak symmetry breaking
appear in various different quantities that can be measured with high
precision at Higgs factories.   First, there is a modification of the
top quark and bottom quark Yukawa couplings. Second, loop effects from
heavy top quark partners shift the Higgs couplings,  an effect most
clearly visible in the couplings to $W$, $Z$, and gluons.  Third,
mixings of the top quark that play a role in the generation of the
Higgs potential will typically modify the top quark electroweak couplings to the
$Z$ and $W$.   A comprehensive exploration for signals of Higgs
compositeness should include measurements of all of these observables.

Models in which electroweak symmetry breaking involves new
interactions
of the top quark must also satisfy another experimental constraint.
Necessarily, from $SU(2)\times U(1)$ symmetry, the left-handed $b$
quark must also have the new interactions.   Then there is a danger
that the $Z$ boson decay width to the $b$ will be altered, in conflict
with the very precise measurement of
\beq
    R_b = \frac{\Gamma(Z\to b\bar b)}{\Gamma(Z\to \text{hadrons})},
\eeq{Rb}
  at the $Z$ pole~\cite{ALEPH:2005ab}.
  A realistic model must avoid this constraint.   There are
  model-building solutions to the problem, for example, the imposition
  of a new custodial symmetry~\cite{Agashe:2006at}.  The models that
  we will study here respect this constraint.

In view of the considerations just described, we will study in this
paper the effects on the Higgs boson and the top quark generated at
1-loop order in the Little Higgs model presented in~\cite{Katz:2003sn}.   This
model is a variant of the model known as the
``Littlest Higgs model''~\cite{Arkani-Hamed:2002ikv}.  The
strong-interaction symmetry breaking pattern is $G/H = SU(5)/SO(5)$.
The Higgs fields are pNGBs in this coset space.  The model contains
only one quark with the quantum numbers of $b_L$, thus eliminating
the possibility of mixing of the $b_L$ with heavier states, the most
important
source of modification of
the $b_L$ coupling to the $Z$.   The price of this is that the model
has a larger than expected set of vectorlike top quark partners, including a partner
with charge $5/3$.  All of these new fermions must have masses above
about 1.2~TeV to avoid constraints from LHC searches.  We choose
this model to illustrate the effects
that these vectorlike fermions can have on the Higgs boson and top
quark properties when these are measured with high precision.    Other
Little Higgs models satisfying the constraints listed above will have
similar effects on the Higgs boson and top quark sectors.

The structure of this paper is as follows:   In Section~II, we 
review the general structure of Little Higgs models and the specific
properties of the $SU(5)/SO(5)$ just described.  We also 
review
the mechanism of collective symmetry breaking and explain how it gives rise to
the
top quark sector of the model.  In Section~III, we analyze how this
structure impacts 
the Higgs Yukawa couplings to fermions and bosons.  In Section IV, we present
the results of a complete scan of the parameter space of the
model.  This will illustrate the mass reach at which the influence
of these heavy fermions can be observed in the precision measurements
of Higgs Yukawa couplngs.  Section V discusses the influence of the
heavy partners on the electroweak couplings of the top quark and
illustrates the mass reach from the precision measurement of these
couplings.  This is a separate set of measurement that provides
additional information on the physics of the vectorlike top quark partners.
Section VI gives  some conclusions.

\section{Electroweak Symmetry Breaking with Higgs as a pNGB}
\label{sec:composite models}

In this section, we review the elements of the Little Higgs model
of Katz et al.~\cite{Katz:2003sn}.  This model is based on a strongly
coupled field theory with the spontaneous symmetry breaking pattern
$SU(5) \to SO(5)$.    We remind the reader that this is the expected
pattern of chiral symmetry breaking for a strongly coupled Yang-Mills
theory with 5 elementary fermions in a real representation of a
confining gauge group. We first give our description of
the Nambu-Goldstone fields that arise from this breaking of the global symmetry.
 We then present the terms involving the fermions of the
model, the top quark and its vectorlike partners.   The mass  terms
and Yukawa
couplings in this sector give the explicit $SU(5)$ symmetry breaking
that will eventually lead to the nonzero Higgs potential and to the
Higgs mass and vacuum expectation value.

\subsection{$SU(5)/SO(5)$ Non-Linear Sigma Model}

It is convenient to write the generators of $SU(5)$ in the {\bf 5}
representation
as $3\times 3$ block matrices with blocks of size $ (2,1,2)$,
corresponding to an $SU(2)\times U(1)\times SU(2)$ subgroup of
$SO( 5) \in SU(5)$. In the model,  The generators of the two $SU(2)$ subgroups are then written
\beq
Q_1^a = \begin{pmatrix}
  \sigma^a/2 &  & \cr  & 0 & \cr  &  & 0\cr
     \end{pmatrix} \
        , \qquad
         Q_2^a = \begin{pmatrix} 0 &  & \cr  & 0 & \cr  &  &
           -\sigma^{aT}/2\cr  \end{pmatrix}  \ .
         \eeq{eq:mySUtwo}
 the  $U(1)$ generator is 
         \beq
         Y = \half \begin{pmatrix} \One & & \cr  & 0 & \cr
              & & -\One\cr  \end{pmatrix} \ ,
         \eeq{eq:myUone}
         where $\One$ is the $2\times 2$ unit matrix.  In the
         model \cite{Katz:2003sn}, the electroweak $SU(2)$ will be
         generated by $Q_1^a + Q_2^a$ and will be an exact symmetry
         at the level of the breaking of $SU(5)\to SO(5)$ symmetry
         breaking. The  orthogonal generators
         $Q_1^a - Q_2^a$ will 
         be spontaneously broken at this level.   From here on, we will refer to
         these as $SU(2)$ and $SU(2)'$, respectively.
       
We describe the breaking $SU(5)\to SO(5)$ by a symmetric $5\times 5$
matrix  whose vacuum expectation value is proportional to 
\beq
\begin{pmatrix}
 0&0&\One\cr
 0&1&0\cr
 \One&0&0\cr
\end{pmatrix}
\eeq{eq:Sigma0}
 After symmetry breaking, the unbroken generators of $SU(5)$ are those
that satisfy
\beq
T^a \Sigma_0 + \Sigma_0 T^{a T} = 0
\eeq{eq:myunbroken}
and the spontaneously broken generators satisfy
\beq
X^i \Sigma_0  - \Sigma_0 X^{iT} = 0 \ . 
\eeq{eq:mybroken}
The generators of $SU(2)$
belong to the unbroken subgroup.

Each NGB is associated with a broken generator.  We can display these
as a matrix  $\Pi = \pi^a X^a$, given explicitly by
\beq
\Pi  =
\begin{pmatrix}
\frac{\eta\, \One}{\sqrt{40}} + \frac{ \theta^a\sigma^a}{\sqrt{8}} & \frac{h^T}{\sqrt{2}} & \phi \cr
\frac{h^\dagger}{\sqrt{2}} & -\frac{2\eta}{\sqrt{10}} & \frac{\tilde h}{\sqrt{2}} \cr
\phi^\dagger & \frac{\tilde h^\dagger}{\sqrt{2}}
& \frac{\eta\, \One}{\sqrt{40}} -\frac { \theta^a\sigma^{aT}}{\sqrt{8}}\cr
\end{pmatrix}
\eeq{eq:myPi}
where $h$ is a complex $SU(2)$ doublet, and $\phi$ is a
complex symmetric $2\times 2$ matrix, that is, a complex $SU(2)$
triplet. In \leqn{eq:myPi},
\beq
h = (h^+, h^0)  \ , \qquad    \tilde h =  (-h^0, h^+)
\eeqn
The field $\eta$ is a singlet that is also an exact NGB.  This must be
given mass, but it will not appear our analysis. If $SU(2)'$ is
gauged, as it will be in this model,  the fields
$\theta^a$ will be eaten in the process of $SU(2)'$ mass generation.

Fluctuations along the broken directions included in \leqn{eq:myPi} are encoded in
$\Pi$ and the non–linear sigma field
\beq
\Sigma(x) = e^{2i\Pi(x)/f}\,\Sigma_0 \  .
\eeq{eq:Sigma}
Here $f$ is the analog of the pion decay constant, with a value in the
multi-TeV range.   The kinetic term for the NGB fields in the
Lagrangian is
\beq
\mathcal{L}_K = \frac{f^2}{4}\,\mathrm{tr}\!\left|D_\mu\Sigma\right|^2.
\eeq{eq:Lkin}
with the coupling to the $SU(2)\times U(1)$ gauge fields
\beqa
D_\mu\Sigma&=&\partial_\mu\Sigma
-  i g_j W_{\mu j}^{a} \big(Q_j^a\Sigma+\Sigma Q_j^{aT}\big)
\CR
&&\hskip 0.5cm
 - i g' B_\mu\big(Y\Sigma+\Sigma Y^T\big)\ .
\eeqa{eq:Dmu}

\subsection{The top sector and the top quark Yukawa coupling}

 In the SM, the top quark is described by a left-handed SU(2) doublet
 field $Q = (t_L, b_L)$ and left-handed singlet field $\bar T = \bar
 t_L$, the antiparticle of the $t_R$.  In Little Higgs models, the
 quadratic divergence in the Higgs mass is cancelled by the
 contributions of vectorlike fermions arising from the strong
 interaction sector.   In this model, we introduce left-handed
 fermions $X$ and $\bar X$ in the ${\bf 5}$  and $\bar {\bf 5}$ of
 SU(5), forming a set of heavy vectorlike quarks.   The  components of
 these multiplets---$(p, \tilde t, \tilde q)$ and $(\bar p$, $\bar
 {\tilde t}, \bar{\tilde q})$, respectively---are listed in
 Table~\ref{tab:VLQcontent}, along with their quantum numbers under
 $SU(2)'\times SU(2)\times U(1)_Y$.

\begin{table}
\begin{center}
\begin{tabular}{|c|c|c|c|c|c|}
\hline
 &  & $SU(3)_c$ & $SU(2)^{\prime}$ & $SU(2)$ & $U(1)_Y$ \\
\hline
\multirow{3}{*}{$X$}
 & $p$           & 3 & $1$ & $2$ & $7/6$ \\
 & $\tilde t$    & $3$ & $1$ & $1$ & $2/3$ \\
 & $\tilde q$    & $3$ & $2$ & $1$ & $1/6$ \\
\hline
\multirow{3}{*}{$\bar X$}
 & $\bar{\tilde q}$ & $3$ & $1$ & $2$ & $-1/6$ \\
 & $\bar{\tilde t}$ & $3$ & $1$ & $1$ & $-2/3$ \\
 & $\bar{p}$        & $3$ & $2$ & $1$ & $-7/6$ \\
\hline
\end{tabular}
\end{center}
\caption{Components of the vectorlike quark multiplets $X$ and $\bar
  X$.} 
\label{tab:VLQcontent}
\end{table}

 These assignments lead to fermion-Higgs Lagrangian terms 
 \beq
\mathcal{L}_t=\lambda_1 f\,\bar X\,\Sigma^\dagger X
+\lambda_2 f\,\bar{\tilde q}\,Q
+\lambda_3 f\,\bar T\,\tilde t+\text{h.c.}
\eeq{eq:Lt-schematic}
These terms produce vectorlike
fermion masses and Yukawa
couplings.  It is convenient to write the mass matrix in the basis
$(\bar p_t,\bar t,\bar{\tilde q}_t,\bar T)\!\times\!(p_t,\tilde
t,\tilde q_t,Q_t)$.  Before electroweak symmetry breaking, with
$\langle h \rangle = 0$, this mass
matrix takes the form
\beq
\begin{pmatrix}   \lambda_1 f  &    0 & 0 & 0 \cr
                             0 & \lambda_1 f  & 0 & 0 \cr
                             0 &  0 & \lambda_1 f  & \lambda_2 f \cr
                             0 & \lambda_3 f  & 0 & 0 \cr
                           \end{pmatrix}   \ .
 \eeq{eq:zeromassmatrix}  
This structure yields 3 massive charge 2/3 quarks and one quark that
remains massless at this level.   This will be the physical top quark,
which can gain mass only from electroweak symmetry breaking.
The components of the physical top quark are given (as
left-handed fermions) by 
\beq
q_3  \equiv \frac{\lambda_2\,\tilde
  q-\lambda_1\,Q}{\sqrt{\lambda_1^2+\lambda_2^2}} \ . 
\eeq{eq:q3}
and
\beq
\bar t \equiv \frac{\lambda_3\,\bar{\tilde t}-\lambda_1\,\bar
  T}{\sqrt{\lambda_1^2+\lambda_3^2}}\ .
\eeq{eq:tbar}
They will obtain mass from a Yukawa coupling
\beq
\lambda_t\,h\,\bar t\,q_3+\text{h.c.}
\eeqn
with 
\beq
\lambda_t = 
\frac{\lambda_1\lambda_2\lambda_3}
{\sqrt{\lambda_1^2+\lambda_2^2}\,\sqrt{\lambda_1^2+\lambda_3^2}}\,.
\eeq{eq:lamt}
After electroweak symmetry breaking, the mass matrix will take
the form shown in  Table~\ref{tab:MassMatrix23}, with
$\theta= \langle h\rangle/(\sqrt{2}\,f)$.

\begin{table}
\begin{center}
\begin{tabular}{|c|c|c|c|c|}
\hline
 & $p_t$ & $\tilde{t}$ & $\tilde{q}_t$ & $Q_t$ \\
\hline
$\bar{p}_t$
  & $\lambda_1 f \cos^2\theta$
  & $\lambda_1 f \dfrac{i}{\sqrt{2}} \sin 2\theta$
  & $-\lambda_1 f \sin^2\theta$
  & $0$ \\
\hline
$\bar{\tilde{t}}$
  & $\lambda_1 f \dfrac{i}{\sqrt{2}} \sin 2\theta$
  & $\lambda_1 f \cos 2\theta$
  & $\lambda_1 f \dfrac{i}{\sqrt{2}} \sin 2\theta$
  & $0$ \\
\hline
$\bar{\tilde{q}}_t$
  & $-\lambda_1 f \sin^2\theta$
  & $\lambda_1 f \dfrac{i}{\sqrt{2}} \sin 2\theta$
  & $\lambda_1 f \cos^2\theta$
  & $\lambda_2 f$ \\
\hline
$\bar{T}$
  & $0$
  & $\lambda_3 f$
  & $0$
  & $0$ \\
\hline
\end{tabular}
\end{center}
\caption{Full mass matrix for the charge 2/3 fermions.  In this table,
  $\theta= \langle h\rangle/(\sqrt{2}\,f)$. }
\label{tab:MassMatrix23}
\end{table}

There are several points to note about this structure of the mass
generation.   First, the $b_L$ remains massless at this stage of the
analysis.  Its mass will be supplied by additional, smaller, Yukawa
terms.   There is no other fermion with the same quantum numbers,
before $SU(2)\times U(1)$ symmetry breaking,  that can mix with it.
Second, there is a vectorlike fermion in the theory with charge 5/3.
Thus the mass of this fermion does depend on $h$, and so it  does not
contribute to the Higgs boson properties through loop
corrections.  Finally, the generation of the  top quark Yukawa coupling
$\lambda_t$ requires all
three coefficients in \leqn{eq:Lt-schematic} to be nonzero.

\subsection{Higgs Potential from Collective Symmetry Breaking}
\label{sec:LH}

To make explicit predictions from this theory, we will need the
detailed formula for the Higgs potential.   This is given by exchanges
of the $SU(2)$ and $SU(2)'$ vector bosons within the strong
interaction sector and loop corrections from the vectorlike top quark
partners.  This discussion follows closely the analysis in \cite{Katz:2003sn}.

In the nonlinear sigma model description, the potential terms from
$SU(2)$ and $SU(2)'$ vector boson
exchange
appear as quadratic divergences cut off at the scale $f$.   These
terms reflect the structure of an $SU(5)$-invariant theory broken in
order $g^2$ by weak gauging.   The unique form is
\beqa
V_{\text{gauge}}(\Sigma)
&=&  C\,g_j^{\,2}\,f^4 \sum_a
\mathrm{tr}\!\Big[(Q_j^{a}\Sigma)(Q_j^{a}\Sigma)^{\!*}\Big] \nonumber
\\
& &\quad + C\,g^{\prime 2}\,f^4\,
\mathrm{tr}\!\Big[(Y\Sigma)(Y\Sigma)^{\!*}\Big] \ .
\eeqa{eq:Vgauge}
We treat this as an order $g^2f^2$ tree-level potential for $h$ and $\phi$.
Expanding to quadratic order in $\phi$ and quartic order in $h$ gives
\beqa
V_{\text{gauge}}
&=&
C\,g_1^2 f^2\,
\Bigg|\,
\phi_{ij}-\frac{i}{2f}(h_i h_j+h_j h_i)
\,\Bigg|^2
\CR
&&
+\;C\,g_2^2 f^2\,
\Bigg|\,
\phi_{ij}+\frac{i}{2f}(h_i h_j+h_j h_i)
\,\Bigg|^2
\CR
&&
+\;C\,g'^2\!\left[
f^2\!\Big(2\,h_i^\dagger h_i+4\,\phi^\ast_{ij}\phi_{ij}\Big)
-\frac{1}{3}\,(h^\dagger h)^2
\right]\ .
\eeqa{eq:Vexpand}
This yields a positive mass for the $\phi$ triplet,
\beq
m_\phi^2\;=\;C\,(g_1^2+g_2^2+4g'^2)\,f^2 \ .
\eeq{eq:mphi}
Notice the the Higgs doublet $h$ receives a mass from \leqn{eq:Vexpand}, but
only from the $U(1)$ boson exchange contribution. This mass term will be a relatively
small
contribution to the eventual full Higgs potential. The $SU(2)$ and $SU(2)'$
exchange terms each leave invariant $SU(3)$ symmetries that forbid a
mass for the PNGBs and so cannot generate Higgs masses at this order.
However, integrating out the $\phi$ fields generates an effective
quartic coupling for the Higgs doublet,
\beqa
\lambda \;=\; C\,
\frac{\;4 g_1^2 g_2^2 + \dfrac{11}{3}(g_1^2+g_2^2)g'^2 - \dfrac{4}{3}g'^4\;}
     {\,g_1^2+g_2^2+4g'^2\,} \ .
\eeqa{eq:lambda}
The 1-loop corrections to the potential \leqn{eq:Vexpand} do generate
Higgs mass terms, since these involve terms that break both $SU(3)$
global symmetries.   These give the  Higgs mass terms
\beq
\delta m_h^2\;=\;\frac{9g^2\,M_{W'}^{\,2}}{64\pi^2}\,
\log\!\frac{\Lambda^2}{M_{W'}^{\,2}}\,.
\eeq{eq:dmh-Wp}
from the $W'$ exchanges, and 
\beq
\delta m_h^2\;=\;\frac{\lambda}{16\pi^2}\,M_\phi^{\,2}\,
\log\!\frac{\Lambda^2}{M_\phi^{\,2}}\,.
\eeq{eq:dmh-phi}
from the massive $\phi$ exchanges.  Both of these effects are of order
$g^4f^2$.  Note that all three Higgs mass terms that we have
identified so far give rise to positve contributions to $\delta
m_h^2$. 

The contribution of the top partners to the Higgs potential is
generated at 1-loop order in the Yukawa couplings in \leqn{eq:Lt-schematic}
by the Coleman-Weinberg formula
\beq
\Delta V_{\text{CW}}^{(f)} \;=\;
-\frac{3}{16\pi^2}\,
\mathrm{tr}\!\Big(M_f(\Sigma) M_f^\dagger(\Sigma)\Big)^2
\log\tfrac{M_f(\Sigma) M_f^\dagger(\Sigma)}{\Lambda^2}\Big.
\eeq{eq:CWfermion}
where $M(\Sigma)$ is the mass matrix given in
Table~\ref{tab:MassMatrix23}.  It is useful to have explicit expressions for
the masses of the 3 heavy top quark partners.  These are:
\beqa
M_1&=& \lambda_1 f \ , \nonumber  \\
M_2& =& \left(
a^2+\frac{\lambda_t^2\,\langle h\rangle^2\,b^2}{a^2-b^2}
-\frac{\lambda_t^4\,\langle h\rangle^4\,(a^4-a^2 b^2+b^4)}{(a^2-b^2)^3}
\right)^{\!\!1/2}\ , \nonumber \\
M_3 & = & \left(
b^2-\frac{\lambda_t^2\,\langle h\rangle^2\,a^2}{a^2-b^2}
+\frac{\lambda_t^4\,\langle h\rangle^4\,(a^4-a^2 b^2+b^4)}{(a^2-b^2)^3}
\right)^{1/2}\, , \nonumber \\
\eeqa{eq:3Ms}
with
\beq
a^2=(\lambda_1^2+\lambda_2^2)f^2 \ ,
\qquad
b^2=(\lambda_1^2+\lambda_3^2)f^2\ .
\eeq{eq:a-b-defs}

For a generic set of Yukawa couplings,
this expression \leqn{eq:CWfermion} would yield a contribution to the
Higgs mass of
order $\lambda^2f^2$, where here $\lambda$ represents a Yukawa coupling.
However, the charge 2/3 mass matrix exhibits
a number of cancellations that are summarized as
\beqa
0 &=& \frac{\partial}{\partial\theta}\,\mathrm{Tr}[M^\dagger M ]\ , \\ 
0 &=&
\frac{\partial}{\partial\theta}\,\mathrm{Tr}\big[(M^\dagger M)^2\big]\ .
\eeqa{eq:traces}
These implement the collective symmetry breaking.   The first of these
cancellations insures that the Higgs mass generated from the
Coleman-Weinberg potential begins only in order $\lambda_i^4 f^2$.
The second cancellation implies that this term is finite and not
enhanced by a factor $\log(f^2/M^2)$.   The explicit expression for the
heavy quark contribution to the Higgs potential is
\beqa
\delta V_{\rm eff}(h)
&=&
-\frac{3\,\lambda_t^2}{8\pi^2}\,(h^\dagger h)\,
\frac{a^2 b^2}{a^2-b^2}\,
\log\!\frac{a^2}{b^2}
\nonumber \\ 
& & \hskip -0.4in
+\frac{3\,\lambda_t^4}{16\pi^2}\,(h^\dagger h)^2
\Bigg[
\frac{(a^2+b^2)}{(a^2-b^2)^3}\nonumber \\
& & \hskip -0.2in \cdot \Big( (3a^4+3b^4-4a^2 b^2)
                      \log\!\frac{a^2}{b^2}
      - (a^4-b^4) \Big) \nonumber \\ 
&& \hskip 0.7in + 2\,\log\!\frac{ab}{h^2}
\Bigg] \ .
\label{eq:Veff-top}
\eeqan

The heavy top partner contribution to $\delta
m_h^2$ is negative.  This is a general feature of Little Higgs models,
that integrating out the heavy quark partners  generates a potential for the
Higgs field with an instability to electroweak symmetry breaking.
According to \leqn{eq:lamt}, the Yukawa couplings in
\leqn{eq:Lt-schematic} will be larger than the top quark Yukawa
coupling $y_t = 1$,  so this term easily overcomes the positive
contribution to  $\delta
m_h^2$ from integrating out the heavy bosons.  The required large size
of the underlying Yukawa couplings will lead to large radiative
corrections to Higgs couplings.  This is possible even while these
Yukawa couplings stay in the region in which they can be treated perturbatively.

\begin{figure*}
\begin{center}
\includegraphics[width = 0.6\linewidth]{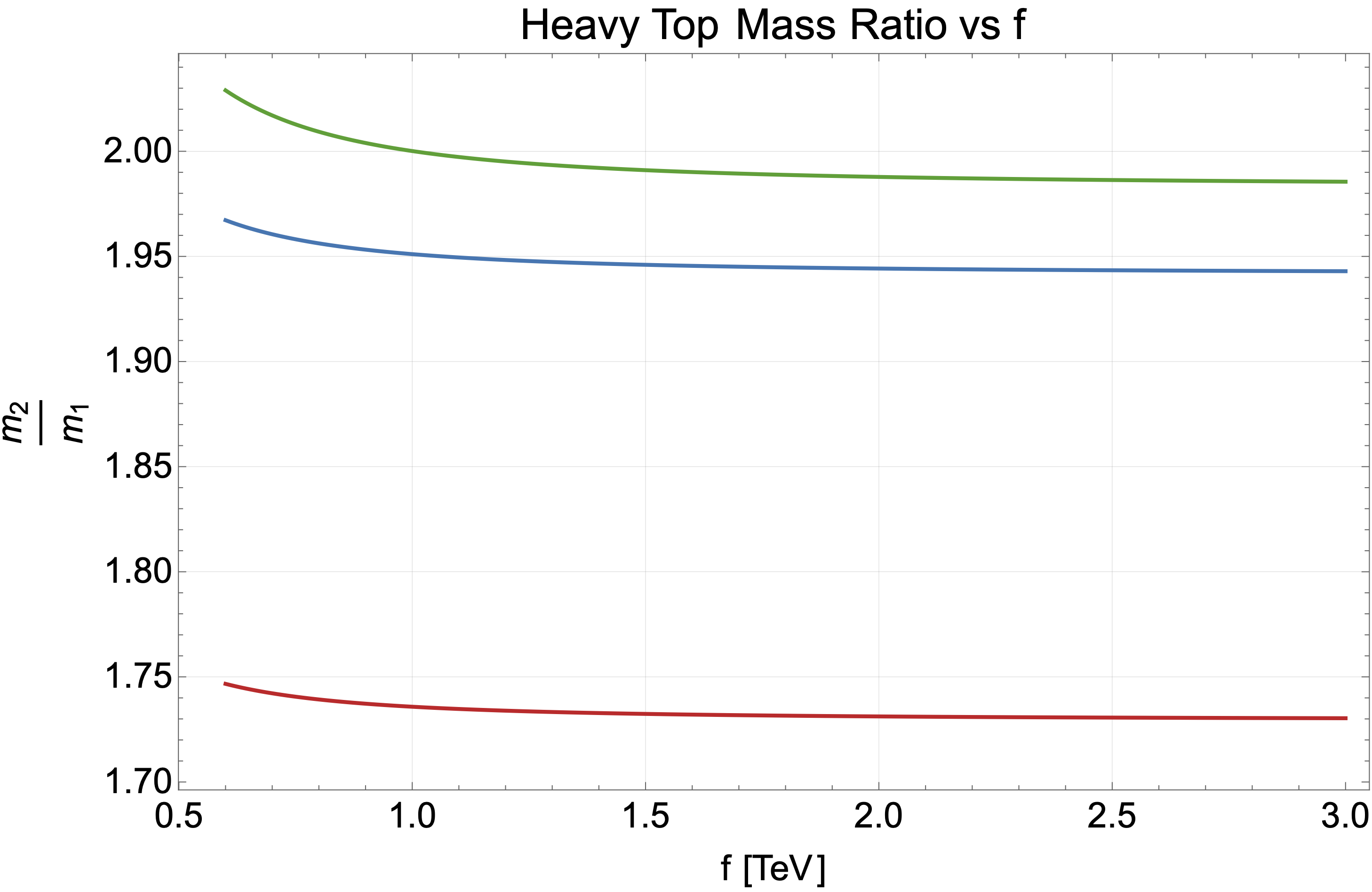}
\end{center}
\caption{The mass ratio of the first and second heavy top partners as a function of $f$. The three curves show the parameter choices, top to bottom:  $\lambda_2/\lambda_3 = 1.3, 1.5, 1.7$.}
\label{fig:1}
\end{figure*}

\begin{figure*}
\begin{center}
\includegraphics[width = 0.6\linewidth]{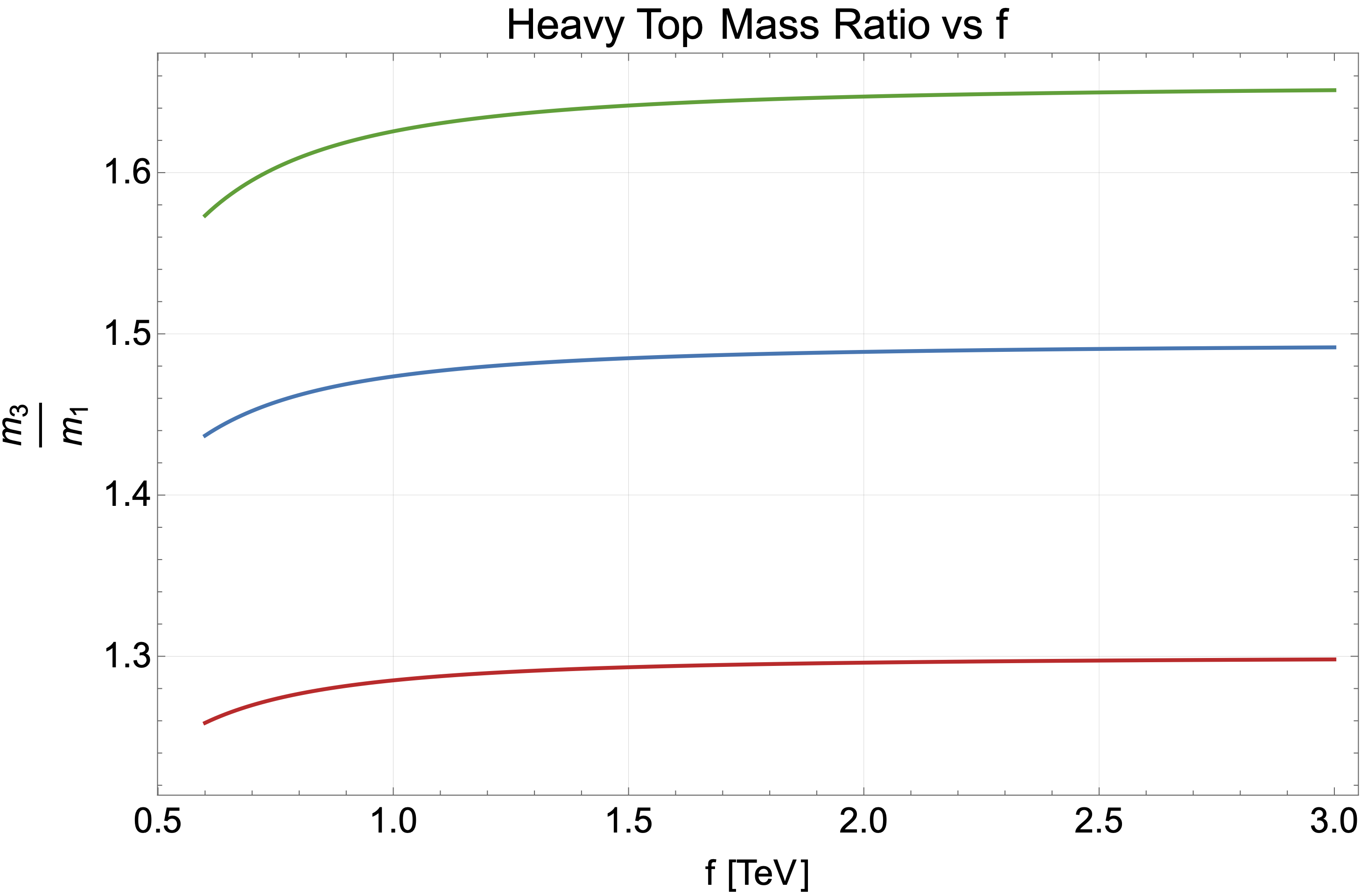}
\end{center}
\caption{The mass ratio of the first and third heavy top partners as a function of $f$.  The three curves show the parameter choices, top to bottom:  $\lambda_2/\lambda_3 = 1.3, 1.5, 1.7$.}
\label{fig:2}
\end{figure*}

\section{Modification of the Higgs Boson Couplings}
\label{sec:Couplings}

By minimizing the potential computed in the previous section, and
constraining the Higgs boson mass, the Higgs field vev, and the top
quark mass to their measured values, we determine the allowed
parameter space for the model \cite{Katz:2003sn}.  Our next task is to
compute the couplings of the Higgs boson over this parameter space
and, in particular, the predicted deviations from their SM values.
The modification of Higgs couplings in Little Higgs models has been
studied some time ago by Han, Logan,
McElrath, and Wang~\cite{Han:2003gf,Han:2003wu,Logan_2004} and by
Killian and Reuter~\cite{Kilian:2003xt}.

In applying their analysis
to this model, we follow the strategy of \cite{Logan_2004}.   In that
paper, the basic input parameters are taken to be the on-shell values
of $G_F$, $m_Z$, $\alpha$, and $m_h$. In this renormalization scheme,
Higgs couplings receive a universal rescaling factor $y_{G_F}$
and individual rescaling factors $y_A$ for the $h\to A\bar A$
coupling.  Then Higgs partial widths modified at the tree level are given
by, for example,
\beq
\Gamma_b/\Gamma_b|_{SM} =   y_{G_F^2}^2 y_b^2 \ .
\eeqn
The partial width for $h\to gg$, which is generated  at the 1-loop
level,  is given by the
expression
\beq
              \Gamma_g/\Gamma_g|_{SM} = y_{G_F^2}^2 \biggl| { \sum_i y_i
                F_{1/2}(\tau_i) \over y_t
                F_{1/2}(\tau_t) }\biggr|^2 \ ,
              \eeqn
 where the index $i$ runs over the top quark and its vectorlike
 partners in the Little Higgs model, and  $\tau = 4 M_i^2/m_h^2$, and
 \beq
F_{1/2}(\tau)=-2\tau\Big[1+(1-\tau)f(\tau)\Big],
\eeq{eq:Fhalf}
with 
\beq
f(\tau)=
\begin{cases}
\arcsin^2(1/\sqrt{\tau}), & \tau\ge 1,\\[2pt]
-\tfrac{1}{4}\left[\log\frac{1+\sqrt{1-\tau}}{1-\sqrt{1-\tau}}-i\pi\right]^2,& \tau<1.
\end{cases}
\eeq{eq:fscalar}

The parameters $y_A$ depend on the Yukawa couplings in
\leqn{eq:Lt-schematic} and also on the $v$ and the value of the triplet
vev through the parameter
\beq
           x  =  \frac{2\lambda_{h\phi h}}{\lambda_{\phi^2}} 
    \eeqn
where $\lambda_{h\phi h}$ is the Higgs-triplet coupling and $\lambda_{\phi^2}$ is the scalar triplet coupling \cite{Han:2003wu}. Points in the parameter space of the model must satisfy $x < 1$ for
stability of the SM electroweak  symmetry breaking
pattern~\cite{Logan_2004}.  In the following, we will follow
\cite{Logan_2004}
 in representing the
couplings $g_1$ and $g_2$ by
\beq
         c = {g_1\over \sqrt{g_1^2 + g_2^2}} \qquad  s= {g_2\over
           \sqrt{g_1^2 + g_2^2}} \ . 
  \eeq{candsdef}

Evaluating the $y_A$ for our model, we find, for the universal
rescaling
\beq
y_{GF}=1+\left(-\frac{5}{24}+\frac{x^2}{8}\right)\frac{v^2}{f^2} \ ,
\eeq{eq:yGF}
for the bottom and top quark Yukawa couplings,
\beqa
 y_b &=& 1+\frac{v^2}{f^2}\!\left(-\frac{2}{3}+\frac{x}{2}-\frac{x^2}{4}\right)\
 , \nonumber \\
 y_t&=&1+\frac{v^2}{f^2}\!\left(-\frac{2}{3}+\frac{x}{2}-\frac{x^2}{4}+\sum_i
   \frac{\lambda_1^2\lambda_i^2}{(\lambda_1^2+\lambda_i^2)^2}\right) ,
\eeqa{eq:ybyt}
where $i = 1,2,3$, and for the heavy top partners
\beq
y_{T_i}=-\,\frac{v^2}{f^2}\,\frac{\lambda_1^2\lambda_i^2}{(\lambda_1^2+\lambda_i^2)^2}\
   .
   \eeq{eq:yTi}
   
The partial width for $h\to WW^*$ is affected not only by the coupling
rescaling
\beq  
y_W = 1+\frac{v^2}{f^2}\!\left(-\frac{1}{6}-\frac{1}{4}(c^2-s^2)^2\right)
\eeq{eq:yW}
but also by shifts in the values of $m_W$, $m_Z$ and
$c_w = \cos\theta_w$  and $s_w = \sin\theta_w$ given by
\beqa
y_{m_W}^2  & = &
1+\frac{v^2}{f^2}\!\left(-\frac{1}{6}-\frac{1}{4}(c^2-s^2)^2 +
  \frac{1}{4}x^2\right) \nonumber \\ 
y_{m_Z}^2 & = & 1
+ \frac{v^2}{f^2}\!\left[-\frac{5}{12} + \frac{1}{2} x^2\right]
+ \frac{m_W^2}{M_{Z_H}^2}\ , \nonumber \\ 
y_{c_w}^2 & = & 1
+ \frac{v^2}{f^2}\,\frac{s_w^2}{c_w^2 - s_w^2}\!\left[-\frac{1}{4} x^2\right]
- \frac{s_w^2}{c_w^2 - s_w^2}\,\frac{m_W^2}{M_{Z_H}^2}\ .
\eeqa{eq:ycw2}
In writing these formulae, we have taken account of the fact that the
model includes only one gauged
$U(1)$ group, $U(1)_Y$. The $M_{Z_H}$ term in the above equations is a consequence of the $SU(2)$ mixing, and having only the one gauged $U(1)_Y$ eliminates any $M_{A_H}$ terms which arise in other equations in Appendix A of \cite{Logan_2004}. The partial width for
$h\to WW^*$ is then given by
\beq
   \Gamma_W/\Gamma_W|_{SM} = y_{G_F^2}^2 { y_{m_W}^2\over y_{m_Z}^2
     y_{c_w}^2} \ .
\eeq{eq:GamW}

These equations enable systematic studies of percent-level deviations
in the Higgs couplings.

\begin{figure*}
\includegraphics[width = .8\linewidth]{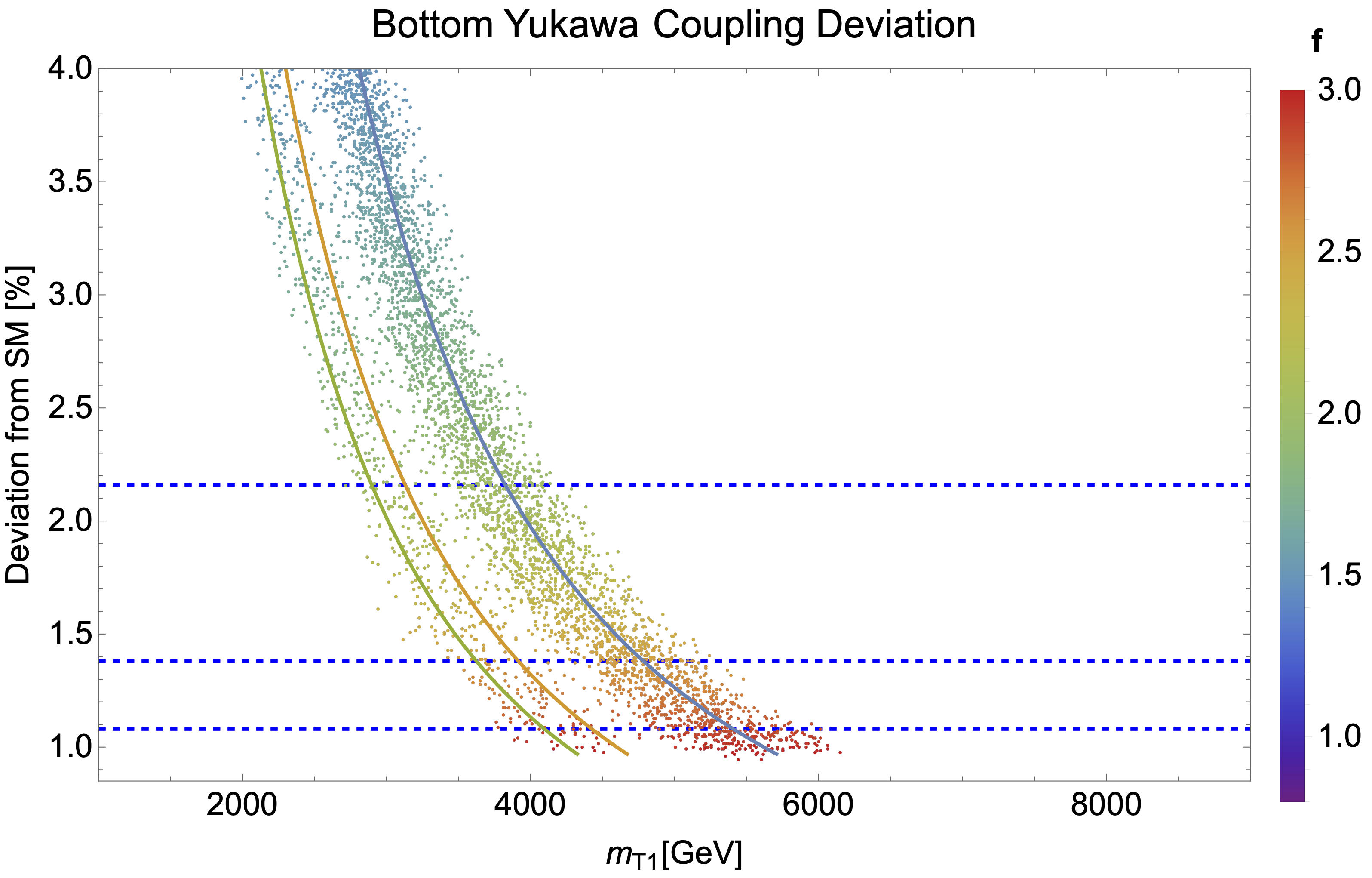}
\caption{Scatter plot depicting the percent change of the bottom quark
  Yukawa coupling from the SM value, over the parameter space of the
  Little Higgs model.  The horizontal axis show the
 mass $M_1$ of the lightest top quark partner.   The solid lines show
 the values in the three scenarios for the ratios of heavy partner
 masses
 $\lambda_2/\lambda_3 = 1.3, 1.5, 1.7$  shown in Figs.~\ref{fig:1} and
 \ref{fig:2}.  The color of each point indicates the value of $f$.
 As in our study~\cite{Maayergi:2025ybi},
   the horizontal dotted lines show the lines of
 3$\sigma$ significance of three stages of the proposed LCF experimental
 program.   The bottom of these  lines is close to the 3$\sigma$
 significance value for FCC-ee and CEPC. }
 \label{fig:3}
\end{figure*}

\begin{figure*}
\includegraphics[width = .8\linewidth]{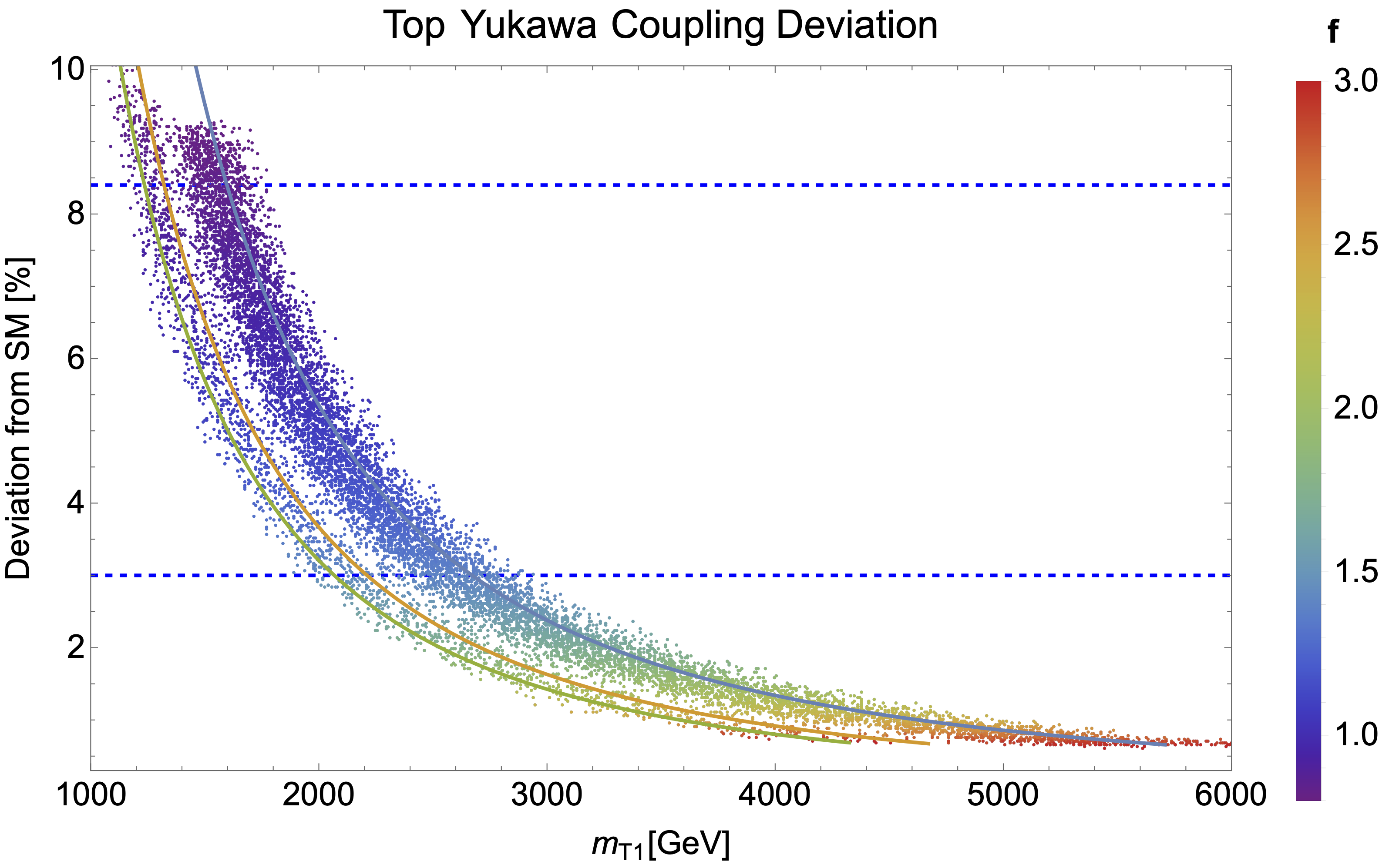}
\caption{Scatter plot depicting the percent change of the top  quark
  Yukawa coupling from the SM value, over the parameter space of the
  Little Higgs model.  The horizontal axis show the
 mass $M_1$ of the lightest top quark partner.  The annotation of the
 plot is as in Fig.~\ref{fig:3}.  In this case, only the lowest two
 horizontal lines are shown. }
\label{fig:t}
\end{figure*}

\begin{figure*}
\includegraphics[width = .8\linewidth]{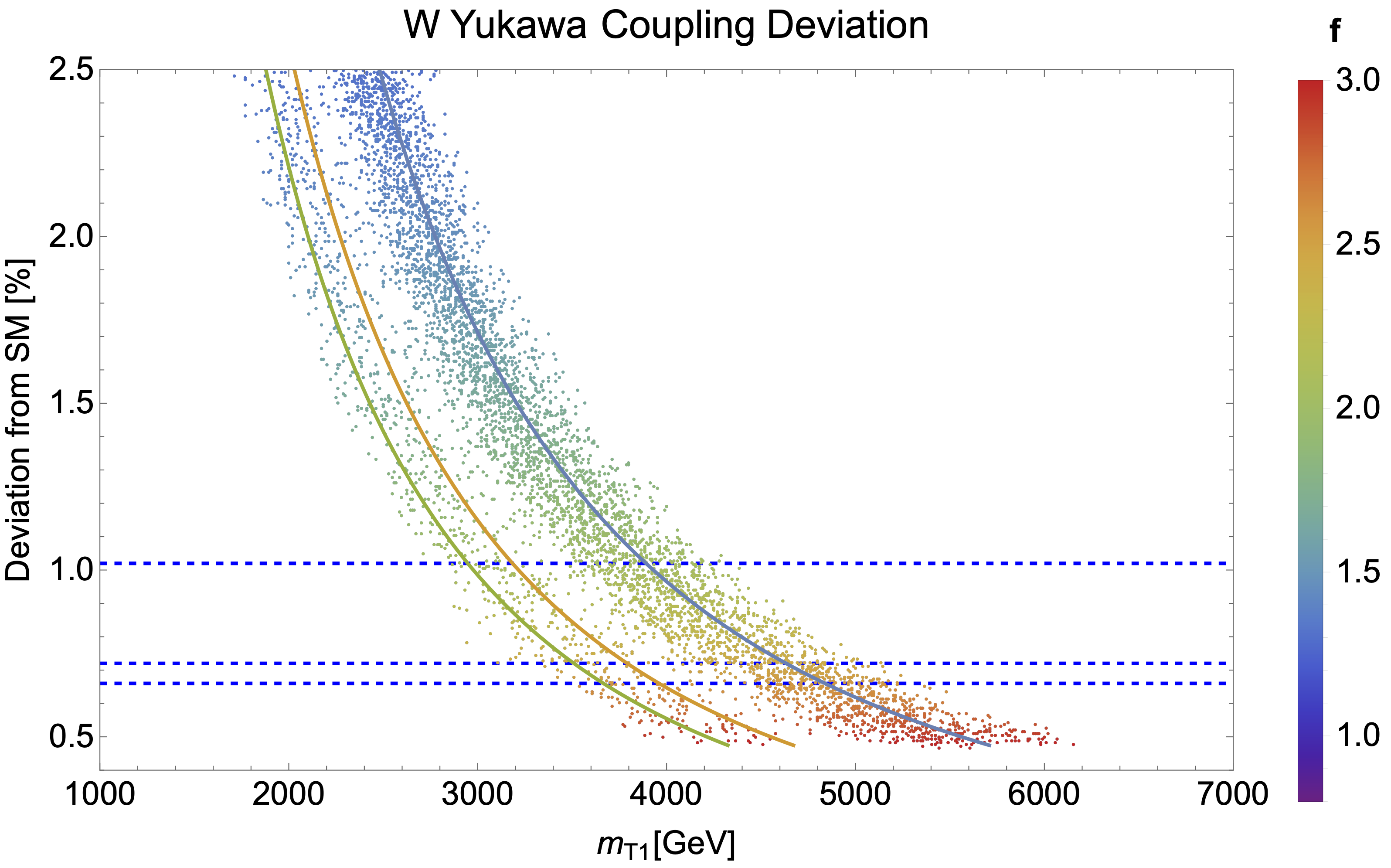}
\caption{Scatter plot depicting the percent change of the Higgs boson
  coupling
  to the $W$ boson from the SM value, over the parameter space of the
  Little Higgs model.  The horizontal axis show the
 mass $M_1$ of the lightest top quark partner.  The annotation of the
 plot is as in Fig.~\ref{fig:3}. }
\label{fig:4}
\end{figure*}

\begin{figure*}
\includegraphics[width = .8\linewidth]{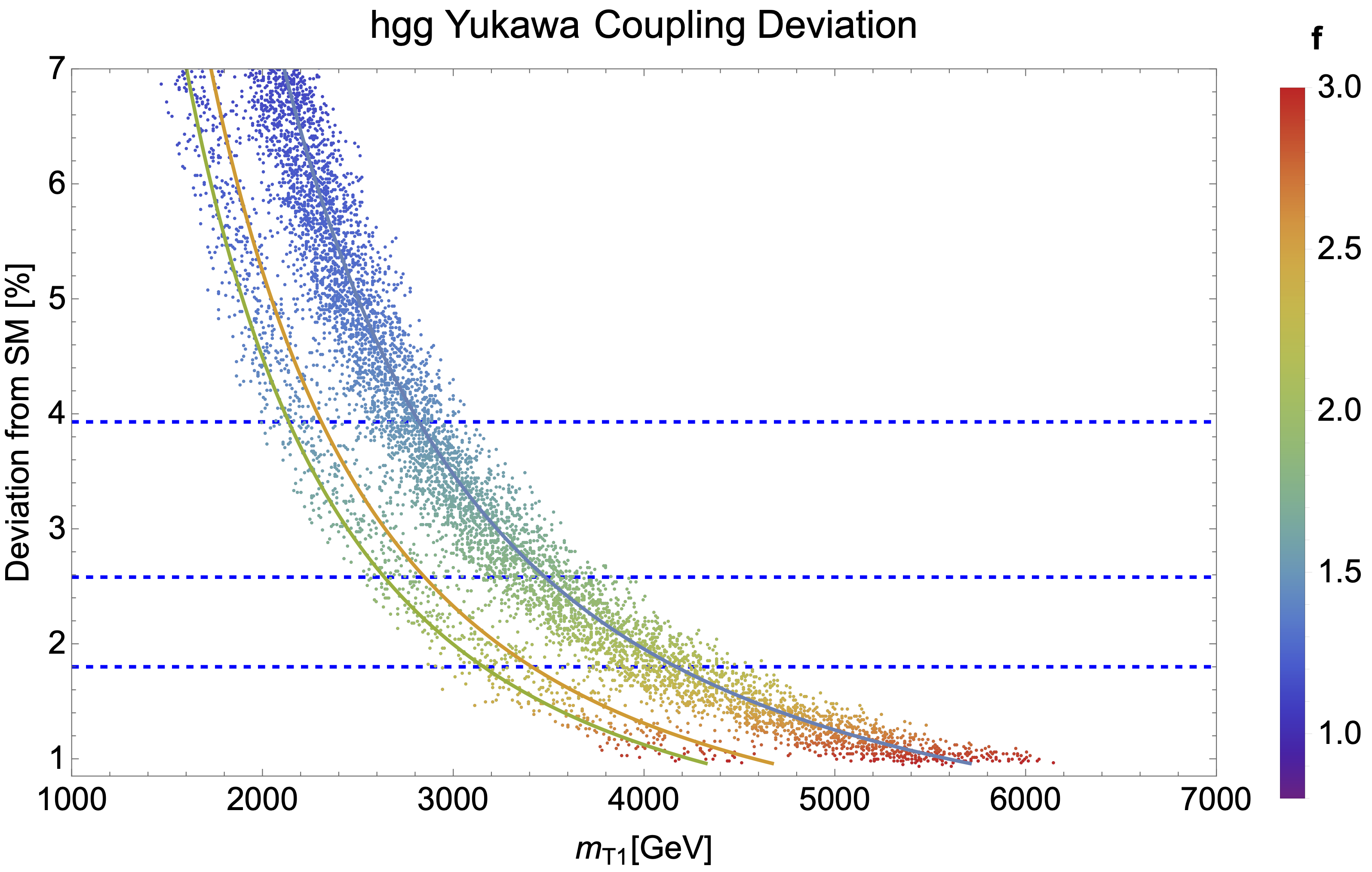}
\caption{Scatter plot depicting the percent change of the Higgs boson
  coupling
  to the gluon from the SM value, over the parameter space of the
  Little Higgs model.  The horizontal axis show the
 mass $M_1$ of the lightest top quark partner.  The annotation of the
 plot is as in Fig.~\ref{fig:3}.}
\label{fig:5}
\end{figure*}

\section{Survey of the Parameter Space}
\label{sec:Couplingscan}

We are now ready to present the main results of this paper, surveying
the  opportunities within the model of \cite{Katz:2003sn} for the
discovery of  deviations from the SM through  high-precision
measurement of the Higgs couplings.

Using the phenomenological formulae given above, we scanned the parameter
space of the model in the following way:  The model depends on the
three coupling constants $g_1$, $g_2$, and $g'$, the three Yukawa
couplings $\lambda_1$, $\lambda_2$, $\lambda_3$, the pion decay
constant $f$, and the 1-loop gauge potential parameter $c$ in
\leqn{eq:Vgauge}.
Fixing the SM input parameters $g$, $g'$, $v$, and
$m_h$ and the top quark mass $m_t$ then gives a 3-dimensional
parameter space.   We explored this space by varying values of the
Yukawa couplings between 0 and 3, values of $f$ up to 3~TeV
(corresponding to strong interaction scales up to $4\pi f = 12$~TeV).
We also scanned over the parameter $C$ between 0.05 and 1 to find a
value which allowed us to reproduce the observed values of the Higgs
mass and vev. The value of the mixing angle $s$ was scanned between
0.05 and 0.5, from which the values of $g_1$ and $g_2$ were
calculated.
These are 
primarily constrained by the
parameter $x$.  For each parameter set, we
minimized the potential given in Sec.~\ref{sec:LH}.  To make this
easier, we first computed $v$ by balancing the fermionic contribution
in \leqn{eq:Veff-top} against the quartic Higgs interaction
\leqn{eq:lambda} and then treating the remaining terms as perturbative
corrections.  We then eliminated solutions violating the stability
condition $x< 1$. 

Using this strategy, it was difficult to find any points that give
reasonable agreement with the known values of $v$ and $m_t$.  To
overcome this difficulty, we carried out an initial random scan of the
parameters using a very large number points.
 After finding a number of seed points that satisfied the
constraints, we then carried out, for each of these, a random scan with
5\% variation of the parameters.   New solutions discovered in this
way were 
used as new seed points, and so, eventually, we discovered extended
parameter regions that satisfy the constraints.   Having accumulated
points in the parameter space in this way, we then used the formula of
Sec.~\ref{sec:Couplings} to compute the expected deviations of the
Higgs couplings.

In the space of solutions that we identified, the top partner masses
generally  have the relation
\beq
M_1 < M_3 < M_2    \ . 
\eeqn
Looking back at the formulae for these masses in \leqn{eq:3Ms}, we see
that the mass $M_1$ is highly correlated with the value of $f$.
However, the solution space seems to divide into two distinct regions
based on the mass ratio of the heavier partners  $M_3/M_2$.   The
dependence of the
mass ratios on these  parameters is shown in Figs.~\ref{fig:1} and
\ref{fig:2}.

Our results for the Higgs coupling deviations are shown in
Figs.~\ref{fig:3}-\ref{fig:5}.   The two distinct regions of parameter
space are clearly indicated.   The upper bound of the allowed regions
have the expected $1/M^2$ decoupling behavior. The solid lines
correspond to the parameter choices for the Yukawa coupling ratio
$\lambda_2/\lambda_3$ shown in Figs.~\ref{fig:1} and \ref{fig:2}.
The horizontal lines indicate the expected 3$\sigma$ significance
expected from the successive stages of the LCF experimental program
with data samples of  3~ab$^{-1}$ at a CM energy of 250~GeV, 
4~ab$^{-1}$ at 550~GeV, and 
8~ab$^{-1}$ at 1000~GeV.   The bottom line is very close to the
3$\sigma$ expectations for the LCF with 4~ab$^{-1}$ at 550~GeV and of
the full FCC-ee and CEPC programs.  For all of these facilities, the
mass reach of the precision Higgs measurements of the individual
Higgs
couplings to $b$, $W$, and $g$
indicates 3$\sigma$ sensitivity to heavy top quark partners
of 3-5 TeV.   The combination of these measurements would give a
discovery of well above 5$\sigma$ significance.   Please note that we
have chosen to display the scan data against the lowest of the three
top partner masses.  There  are models shown for which the value of $M_2$
to which these measurements are sensitive is  greater than  10~TeV.

The measurement expectation for the top quark Yukawa coupling
for LCF at 1000~GeV is a precision of
1\%.  This is essentially the same as the expectation for FCC-hh.
This measurement does not seem to add much significance to the
discovery in any model in this class.


\begin{figure*}[]
\includegraphics[width = 0.8\linewidth]{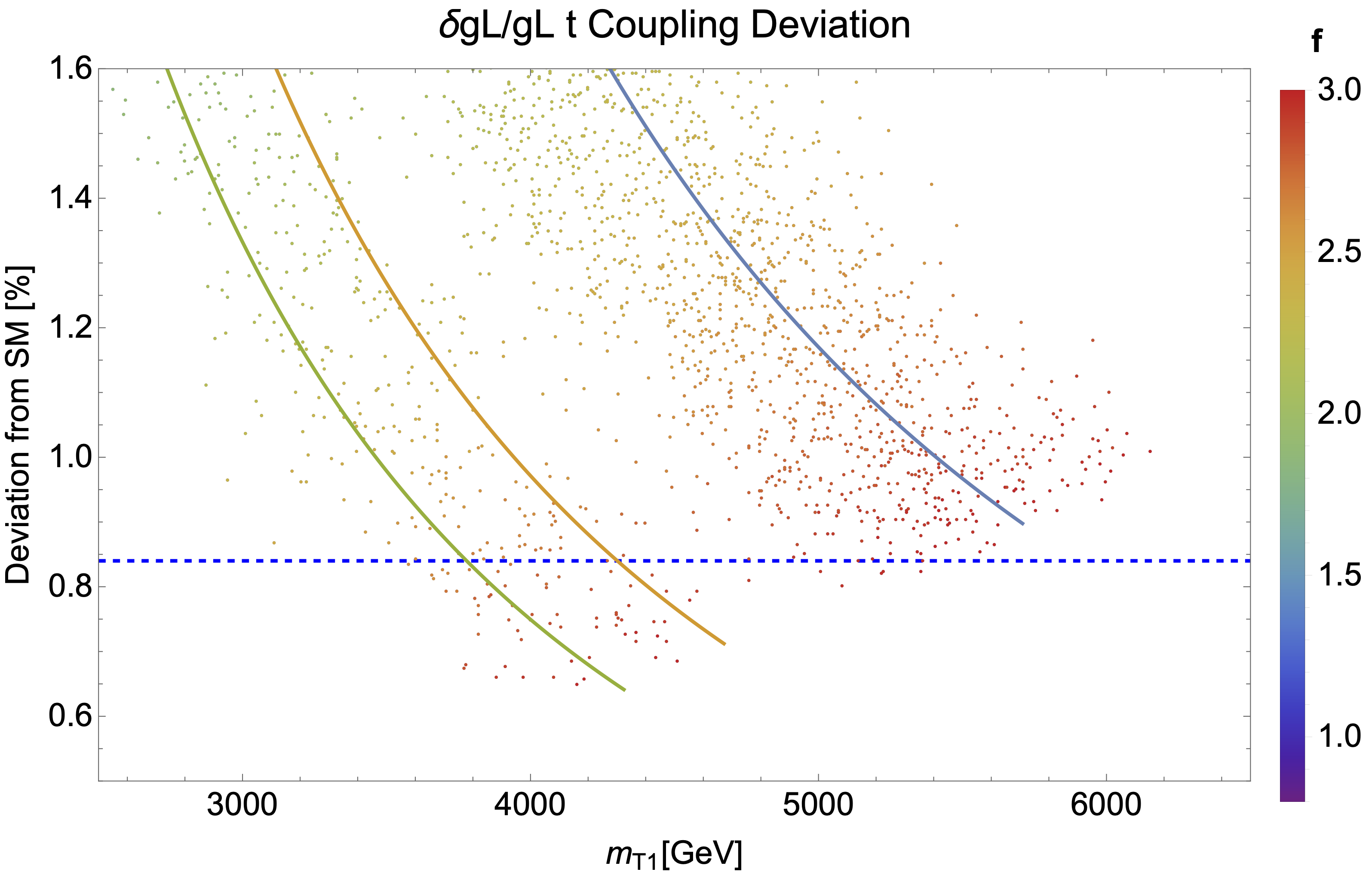}
\caption{Scatter plot depicting the percent change of the  electroweak
  coupling of the $t_L$ to the $Z$ boson over the parameter space of the
  Little Higgs model.  The horizontal axis show the
 mass $M_1$ of the lightest top quark partner.  The annotation of the
 plot is as in Fig.~\ref{fig:3}.   The single horizontal line is the
 line of 3$\sigma$ significance from the LCF program at 550~GeV.}
\label{fig:7}
\end{figure*}

\section{Modification of the Top Quark Electroweak Couplings}
\label{sec:TopCouplings}

We noted in the Introduction that, in addition to modifications of the
Higgs couplings, Little Higgs models will also modify the top quark
coupling to the $Z$ boson.  Using the same scan
that we have presented in the previous section, we can explore the
size of this effect in the parameter space of the model
\cite{Katz:2003sn}.

The electroweak coupling modifications were also worked out in the
general analysis presented in \cite{Han:2003wu}.   Specializing the
formulae presented there to this model, we find for the left- and
right-handed couplings of the $b$ quark to the $Z$ boson
\beqa
 g_L^b &=& {g\over c_w} \biggl[ (-\frac{1}{2}+ \frac{1}{3}s_w^2)  -
 {v^2\over f^2}{c\over 2 s}c_w x_Z^{W'} \biggr] \nonumber \\
 g_R^b &=& {g\over c_w} \biggl[ (+ \frac{1}{3}s_w^2) \biggr]  \ ,
 \eeqa{Zbcouplings}
 and for the left- and right-handed couplings of the top quark
\beqa
 g_L^t &=& {g\over c_w} \biggl[ (+ \frac{1}{2}- \frac{2}{3}s_w^2)  -
 {v^2\over f^2}({x_{L1}^2\over 2} +{x_{L2}^2\over 2}  -
         {c\over 2 s}c_w x_Z^{W'} )\biggr] \nonumber \\
 g_R^b &=& {g\over c_w} \biggl[ (-\frac{2}{3}s_w^2) \biggr]  \ ,
 \eeqa{Zycouplings}
 where
 \beq
x_{L1} =
\frac{\lambda_1^2}{(\lambda_1^2+\lambda_2^2)} \qquad
x_{L2} =\frac{\lambda_1^2 }{(\lambda_1^2+\lambda_3^2)}
\ ,
\eeq{eq:xL}
 and $x^{W'}_Z$ is a correction due to the mixing of the two sets of
 $SU(2)$    bosons, given by
 \beq
 {c\over 2 s}c_w x_Z^{W'} = - \frac{c^2 (c^2-s^2)}{4} \ .
\eeq{eq:xZWprime}
This correction, equal and opposite for up and down quarks, is also
present in the light quark couplings to the $Z$.   The resulting shift
of the $b_L$ coupling then largely cancels
out in the ratio $R_b$. The formulae
of \cite{Han:2003wu} also include larger corrections due to the
mixing of  the $SU(2)$ with heavy $U(1)$ bosons that are not present
in the model that we consider here.

To compare to the capabilities of Higgs factories, we observe that our
Little Higgs model gives no correction to the $t_R$ coupling, while
for the $t_L$ coupling there is a fractional shift
\beq
\frac{\delta g_{L}^{t}}{g_{L}^{t}}
= \frac{v^2}{f^2}\,
\frac{\,x_{L1}^{2}+x_{L2}^{2}
- c_w\,x_Z^{W'}\,\dfrac{c}{s}
}
{1-(4/3) s_w^{2}}\ .
\eeq{eq:dgtL}
The values of this quantity over the parameter space of the model are
shown in Fig.~\ref{fig:7}.   The measurement of this quantity from
$\ee\to t\bar t$ is discussed in \cite{LinearColliderVision:2025hlt},
which expects the precision
\beq
    \frac{\delta g_L^t}{g_L^t}=0.28\%  \ ,
    \eeq{LCFgL}
from the LCF program at 550~GeV.   The excellent sensitivity results
from the fact that the cross section is dominated by s-channel
$\gamma$ and $Z$ exchange.   The two contributing amplitudes can be
separated using beam polarization.

\section{Conclusions}

In this paper, we have displayed the sensitivity of precision
measurements at Higgs factories to an explicit model in which the
Higgs boson is composite, realized as a Goldstone boson of a symmetry
breaking at a high mass scale.  Models of this type obey many
constraints, from precision electroweak measurements and from direct
searches for top quark partners and other heavy states that they
include.  Still, these models provide large parameter spaces in which
ideas about the phenomenology of these models can be tested.  They are
especially interesting as examples of models with large little
hierarchies that appear naturally and point to physics goals beyond the
reach of the LHC.

We have seen here that $\ee$ Higgs factories allow the observation of
many signatures of these models, in the Higgs Yukawa couplings, in other
Higgs couplings that  the heavy particles of the theory  influence through loop effects,
and in the modification of the electroweak interactions of the top
quark.  This is only one of the many Beyond-Standard Model scenarios
for which Higgs factories offer important opportunities for
discovery.

\begin{acknowledgments}
  We continue to appreciate Ann Nelson's insights into the theory of
  electroweak symmetry breaking.
  We are grateful to Roman P\"oschl and Marcel Vos for discussions of
  the measurement of top quark properties at Higgs factories.
The work of KM and DGEW was supported in part by the grant NSF
OIA-2033382.   The work of MEP was supported by the US Department
of Energy under contract DE–AC02–76SF00515.  
\end{acknowledgments}

\bibliography{Bibliography2}

\end{document}